\documentclass[preprint,aps,a4paper,showpacs,amsfonts,amsmath,amssymb]{revtex4}
\usepackage{graphicx}
\newcommand{\sgn}{{\rm sgn}}
\newcommand{\y}{\xi}

\begin{document}

\title{Non-independent continuous time random walks}
\author{Miquel Montero}\email{miquel.montero@ub.edu} 
\author{Jaume Masoliver}\email{jaume.masoliver@ub.edu}
\affiliation{Departament de F\'{\i}sica Fonamental, Universitat de
Barcelona.\\ Diagonal 647, E-08028 Barcelona, Spain}

\begin{abstract}

The usual development of the continuous time random walk (CTRW) assumes that jumps and time intervals are a two-dimensional set of independent and identically distributed random variables. In this paper we address the theoretical setting of non-independent CTRW's where consecutive jumps and/or time intervals are correlated. An exact solution to the problem is obtained for the special but relevant case in which the correlation solely depends on the signs of consecutive jumps. Even in this simple case some interesting features arise such as transitions from unimodal to bimodal distributions due to correlation. We also develop the necessary analytical techniques and approximations to handle more general situations that can appear in practice.

\end{abstract}

\date{\today}
\maketitle

\section{Introduction}
\label{sec1}

For more than four decades, since their introduction in 1965 by Montroll and Weiss \cite{montrollweiss}, continuous time random walks (CTRW's) have been applied to virtually any field in which one wishes to provide a dynamical description on the microstructure of a given random system. A huge number of examples and applications can be found in the literature, of which we only cite a handful: transport in disordered media \cite{montroll2,weissllibre}, random networks \cite{berkowitz}, self-organized criticality \cite{boguna}, electron tunneling \cite{gudowska}, earthquake modeling \cite{sornette,AC06}, hydrology \cite{berkowitz2,dentz}, time-series analysis \cite{grigo2,kutner1} and finance \cite{scalas,raberto,scalas-pre,scalas2,kutner,mmw,rr04,jebo,mmpw}. 

The CTRW generalizes the ordinary random walk since in the latter the steps of the random walker are made at equal intervals of time, while in the CTRW the interval between steps is a continuous random variable. In this sense CTRW is related to several other extensions of random walks in continuous time, like semi-Markov processes or Markov renewal processes~\cite{CM65}, although the seeds of this idea can be traced back to the 1920's with the pure birth Poisson process~\cite{GUY24,WF50}. 

A great number of developments of the CTRW are based on the assumption that the magnitude of the steps (or jumps) and the time intervals between them (also called sojourns) are a two-dimensional set of independent and identically distributed random variables. While in many cases this is a convenient assumption which allows for simple developments, there are some other cases in which independent walks are clearly insufficient to explain some aspects of the physical reality, and correlations between consecutive step sizes and/or waiting times must be considered. We have met with such a case in our study of financial time series regarding their extreme time statistics \cite{montero_lillo} where jump magnitudes (transaction-to-transaction returns) show short-range memory, but it seems to be also relevant in earthquake modelling, where some evidences pointing to the presence of cross-correlations between simultaneous and sequential earthquake magnitudes and recurrence times have been reported~\cite{AC06b}. Successive recurrence times appear to be positively self-correlated as well but, oddly enough, consecutive magnitudes seems to be independent in this case.  

Our goal in this paper is to address the theoretical setting of non-independent CTRW's, a class of random walks that can account for many physical situations. In this kind of walks jumps and sojourns are no longer independent random variables and their value at a given step may depend on previous steps. We develop the formalism for the Markovian case in which the magnitudes of a given jump and time interval only depend on the preceding step. We also obtain a complete solution to the problem when the correlation between consecutive steps solely depend on the sign of the previous jump (that is, whether the previous jumps is increasing or decreasing but not on its magnitude). Particular examples  corresponding to this solvable case allow us to visualize and quantify some interesting consequences of the existence of correlations between steps as, for instance, the transitions from unimodal to bimodal distributions.
Finally, we also develop the necessary perturbation techniques to deal with more general situations. 

The paper is organized as follows: in Sect. \ref{sec2} we outline the traditional CTRW based on the assumption of independence between events. In Sect. \ref{sec3} we present the general setting for non-independent CTRW's that are still amenable to analytical treatment. In Sect. \ref{sec3b} we present an exact solution to the problem. Section \ref{sec4} is devoted to consider weak dependent models for which we develop a complete perturbation technique. Conclusions are drawn in Sect. \ref{sec5}. Although this work is essentially technical some even more technical aspects are in an Appendix.

\section{The independent CTRW}
\label{sec2}

Suppose that a given random process $X(t)$ evolves following a CTRW. In this picture any realization of $X(t)$ consists of a series of step functions and $X(t)$ changes at random times $\cdots,t_{-2},t_{-1}, t_0, t_1, t_2,\cdots$ while it remains fixed in place between successive steps. The interval between these successive steps is a random variable $\Delta t_n=t_n-t_{n-1}$ which we call sojourn or waiting time. At the conclusion of the $n$th sojourn $X(t)$ experiences a random change, or jump, given by 
$$
\Delta X_n=\Delta X_n(\Delta t_n)=X(t_n)-X(t_{n-1})=X_n-X_{n-1}.
$$
In the usual model of the CTRW waiting times $\Delta t_n$ and random jumps $\Delta X_n$ constitute a two-dimensional set of identically distributed random variables, $(\Delta X_n,\Delta t_n)$, described by the corresponding joint probability density function (pdf) $\rho(\xi,\tau)$,
\begin{equation*}
\rho(\xi,\tau)d\xi d\tau={\rm Prob}\{\xi<\Delta X_n\leq \xi+d\xi; \tau<\Delta t_n\leq \tau+d\tau\}. 
\end{equation*}
As usual, two marginal pdf's can be derived from $\rho(\xi,\tau)$: 
\begin{equation}
h(\xi)d\xi={\rm Prob}\{\xi<\Delta X_n\leq \xi+d\xi\},
\label{h}
\end{equation}
and
\begin{equation*}
\psi(\tau)d\tau={\rm Prob}\{\tau<\Delta t_n\leq \tau+d\tau\},
\end{equation*}
just by integrating over the opposite variable:
\begin{equation*}
h(\xi)=\int_{0}^\infty\rho(\xi,\tau)d\tau, \qquad \psi(\tau)=\int_{-\infty}^\infty\rho(\xi,\tau)d\xi.
\end{equation*}
In this set-up, each pair of random variables $(\Delta X_n,\Delta t_n)$ is independent of any other pair $(\Delta X_m,\Delta t_m)$, $m \neq n$, but it is still allowed any degree of correlation between $\Delta X_n$ and $\Delta t_n$ themselves~\footnote{From a mathematical point of view, this implies that $X(t)=X_{N(t)}$, $N(t)=\sup\{n|t_n \leq t\}$, can be thought as a generalization of a semi-Markov process, since $X_n$ is, in general, not a Markov chain but a Markov discrete process with continuous states, and $\Delta t_n$ is a random variable depending on both $X_n$ and $X_{n-1}$ through $\Delta X_n$.}. Some CTRW processes belonging to this category can be found in~\cite{kutner,mmw,rr04,jebo,mmpw} for instance, and sometimes are also named as {\it non-independent\/} models~\cite{rr04}, since the label {\it independent\/} is then reserved for the particular case in which $\Delta X_n$ and $\Delta t_n$ are also {\it mutually\/} independent random variables, {\it i.e.\/} when
\begin{equation}
\rho(\xi,\tau)=h(\xi)\psi(\tau). 
\label{rho_full_indep}
\end{equation}
This divergence in the existing notation may therefore induce to misinterpretation. We prefer to call Eq.~(\ref{rho_full_indep}) the {\it fully-independent\/} CTRW, and keep the term independent CTRW for any memoryless process with arbitrary joint pdf. 

The chief objective of the CTRW formalism is to obtain the probability density function of $X(t)$. This pdf, called the propagator, is defined by
\begin{equation*}
p(x,t)dx={\rm Prob}\{x<X(t)\leq x+dx|X(t_0)=0\},
\end{equation*}
where in what follows we shall assume that the initial jump occurred at $t_0=0$. As is well known the propagator obeys the following renewal equation \cite{mmw}
\begin{equation}
p(x,t)=p_0(x,t)+\int_{0}^{t}dt'\int_{-\infty}^\infty\rho(x-x',t-t')p(x',t')dx'.
\label{renewal_0}
\end{equation}
The function $p_0(x,t)$ is the propagator prior the first jump and, since  the trajectories of $X(t)$ consist in series of step functions, we write
\begin{equation}
p_0(x,t)=\Psi(t)\delta(x),
\label{p0_0}
\end{equation}
where $\Psi(t)$ is the probability that no transaction has occurred before time $t$
\begin{equation}
\Psi(t)=\int_t^\infty\psi(t')dt'.
\label{cumulative}
\end{equation}

We can solve Eq. (\ref{renewal_0}) in terms of the joint Fourier-Laplace transform:
$$
\tilde{p}(\omega,s)=\int_{0}^\infty dt e^{-st}\int_{-\infty}^\infty e^{i\omega t} p(x,t) dx.
$$
The solution is 
\begin{equation}
\tilde{p}(\omega,s)=\frac{\tilde{p}_0(\omega,s)}{1-\tilde{\rho}(\omega,s)},
\label{solution_0}
\end{equation}
where $\tilde{p}_0(\omega,s)$ and $\tilde{\rho}(\omega,s)$ are the joint Fourier-Laplace transforms of the functions $p_0(x,t)$ and $\rho(x,t)$. We easily see from Eqs. (\ref{p0_0})-(\ref{cumulative}) that the explicit form of $\tilde{p}_0(\omega,s)$ is
\begin{equation*}
\tilde{p}_0(\omega,s)=\frac{1-\hat{\psi}(s)}{s},
\end{equation*}
where $\hat{\psi}(s)$ is the Laplace transform of the pausing-time density $\psi(\tau)$.

\section{A general non-independent walk}
\label{sec3}

As we have explained, the CTRW outlined in the preceding section relies on the assumption that the pairs $(\Delta X_n,\Delta t_n)$ are independent two-dimensional random variables.  However, as we have mentioned in Sect. \ref{sec1} there are many situations in which the assumption of independence may be doubtful~\cite{montero_lillo,AC06b}. We shall thus generalize the CTRW formalism in order to account for memory effects due to correlations between different sojourns and/or jump increments. 

Among the many ways of doing this extension we choose a simple, but yet general, method which consists in assuming that the joint density of $k$ consecutive changes, $\rho(\xi_n,\tau_n;\xi_{n-1},\tau_{n-1};\cdots;\xi_{n-k},\tau_{n-k})$, fulfills the Markov property:
\begin{equation*}
\rho(\xi_n,\tau_n;\xi_{n-1},\tau_{n-1};\cdots;\xi_{n-k},\tau_{n-k})
=\prod_{m=0}^{k-1} \rho(\xi_{n-m},\tau_{n-m}|\xi_{n-m-1},\tau_{n-m-1})\cdot\rho(\xi_{n-k},\tau_{n-k}),
\end{equation*}
where
\begin{eqnarray}
\rho(\xi',\tau'|\y,\tau)d\xi'd\tau'&=&{\rm Prob}\{\xi'<\Delta X_n\leq \xi'+d\xi';\nonumber \\
&\quad & \tau'<\Delta t_n\leq \tau'+d\tau'|\Delta X_{n-1}=\y;\Delta t_{n-1}=\tau\}.
\label{rho_mark}
\end{eqnarray}
This means that the pair $(\Delta X_n,\Delta t_n)$ depends on all the previous transitions only through the immediately preceding one $(\Delta X_{n-1},\Delta t_{n-1})$. In an analogous way to the independent walk described by Eq.~(\ref{renewal_0}), the integral equation governing the evolution of the return pdf is given in this case by the renewal equation
\begin{eqnarray}
p(x,t|\y,\tau)&=&p_0(x,t|\y,\tau) \nonumber \\
&+& \int_{-\infty}^{\infty}dx'\int_0^t \rho(x',t'|\y,\tau) p(x-x',t-t'|x',t')dt',
\label{NMrenewal}
\end{eqnarray}
where $p_0(x,t|\y,\tau)$ is the propagator prior the first jump and similarly to Eq. (\ref{p0_0}) we write
\begin{equation}
p_0(x,t|\y,\tau)=\delta(x)\Psi(t|\y,\tau),
\label{NMp0}
\end{equation}
where $\Psi(t|\y,\tau)$ is the cumulative distribution of the waiting time and it is related to the transition density $\rho(x',t'|\y,\tau)$ by
\begin{equation}
\Psi(t|\y,\tau)=\int_t^\infty dt'\int_{-\infty}^\infty \rho(x',t'|\y,\tau) dx'.
\label{NMPsi}
\end{equation}

Observe that in this case the process $X(t)$ is not Markovian because its pdf $p(x,t|\y,\tau)$ depends on both the magnitude of the previous jump $\y=X_0-X_{-1}$ and its sojourn time $\tau=t_0-t_{-1}$; in other words, the probability distribution of the process at a given time depends on two previous times $t_0$ and $t_{-1}$. We should note at this point that even the independent CTRW is, in general, a non-Markovian process. The only case in which the independent CTRW is Markovian is when it is a fully-independent CTRW, and the set of random times $\cdots,t_0,t_1,t_2,\cdots$ is Poissonian, that is, when the pausing time density $\psi(\tau)$ obeys the exponential law \cite{weissllibre}:
\begin{equation*}
\psi(\tau)=\lambda e^{-\lambda \tau}, \qquad (\lambda>0).
\end{equation*}
Let us remark that the dependent CTRW outlined above is always non-Markovian even for this Poissonian density.

We also note that in the case of independent increments discussed in Sect. \ref{sec2} we have 
$$
\rho(x',t'|\y,\tau)=\rho(x',t'),
$$
and Eq. (\ref{NMrenewal}) reduces to Eq. (\ref{renewal_0}). 

The integral equation given in Eq. (\ref{NMrenewal}) is the general equation that governs the evolution of the random process $X(t)$ and it must be solved if we want to obtain the propagator for this non-independent case. Contrary to the independent case, Eq. (\ref{NMrenewal}) cannot be solved, for any form of the joint density $\rho$ by means of transform methods. Indeed, the Laplace transform with respect to $t$ of Eq. (\ref{NMrenewal}) is
\begin{eqnarray*}
\hat{p}(x,s|\y,\tau)&=&\hat{p}_0(x,s|\y,\tau) \nonumber \\
&+& \int_0^\infty dt e^{-st}\int_0^t dt'\int_{-\infty}^{\infty}\rho(x',t'|\y,\tau) p(x-x',t-t'|x',t')dx',
\end{eqnarray*}
where the hat over $p$ and $p_0$ denotes the time Laplace transform. Note that
$$
\int_0^\infty dt e^{-st}\int_0^t dt'\cdots=\int_0^\infty dt'e^{-st'}\int_0^\infty dt'' e^{-st''}\cdots,
$$
where we have exchanged the order of integration and performed the change of variables $t''=t-t'$. Then 
\begin{eqnarray*}
\hat{p}(x,s|\y,\tau)&=&\hat{p}_0(x,s|\y,\tau) \nonumber \\
&+&\int_0^\infty dt' e^{-st'} \int_{-\infty}^{\infty}\rho(x',t'|\y,\tau) \hat{p}(x-x',s|x',t')dx'.
\end{eqnarray*}
Finally, the Fourier transform with respect to $x$ of this equation yields
\begin{eqnarray}
\tilde{p}(\omega,s|\y,\tau)&=&\tilde{p}_0(\omega,s|\y,\tau) \nonumber \\
&+&\int_0^\infty dt' e^{-st'} \int_{-\infty}^{\infty} e^{i\omega x'}\rho(x',t'|\y,\tau) \tilde{p}(\omega,s|x',t')dx',
\label{NMrenewal_2}
\end{eqnarray}
which is the farthest we can go without specifying $\rho(x',t'|\y,\tau)$.

\section{A solvable case}
\label{sec3b}

The integral equation (\ref{NMrenewal_2}) cannot be solved for any arbitrary form of $\rho$. However, for the independent case in which $\rho(x',t'|\y,\tau)=\rho(x',t')$ and $\tilde{p}(\omega,s|x',t')=\tilde{p}(\omega,s)$ we recover from Eq. (\ref{NMrenewal_2}) the solution given by Eq. (\ref{solution_0}). Another case in which the level of difficulty is somewhat reduced is when waiting times and jumps are not related to each other. In such a case the joint density factorizes as:
\begin{equation*}
\rho(x',t'|\y,\tau)=\psi(t'|\y,\tau)h(x'|\y,\tau).
\end{equation*}
We will also assume the further simplification
\begin{equation}
\rho(x',t'|\y,\tau)=\psi(t')h(x'|\y),
\label{independent_0}
\end{equation}
in which correlations between consecutive waiting times have been neglected and we have also assumed that the waiting time pdf, $\psi(t')$, does not depend on the magnitude of the jumps; a situation that, as mentioned above, has been detected in some financial time series \cite{montero_lillo}. 

The renewal equation for the propagator of $X(t)$ now reduces to (cf. Eqs. (\ref{NMrenewal})-(\ref{NMp0}))
\begin{equation*}
p(x,t|\y)=\delta(x)\Psi(t)
+\int_0^t dt' \psi(t') \int_{-\infty}^{\infty}h(x'|\y) p(x-x',t-t'|x',t')dx'.
\end{equation*}
The (time) Laplace transform of this equation yields
\begin{equation}
\hat{p}(x,s|\y)=\delta(x)\hat{\Psi}(s)+
\hat{\psi}(s)\int_{-\infty}^{\infty}h(x'|\y)\hat{p}(x-x',s|x')dx',
\label{NMrenewal_3b}
\end{equation}
where, in terms of the density $\hat{\psi}(s)$, the waiting time distribution function $\hat{\Psi}(s)$ can be written as
$$
\hat{\Psi}(s)=\frac{1-\hat{\psi}(s)}{s}.
$$

Let us now suppose that the conditional jump density $h(x'|\y)$ has the form
\begin{equation}
h(x'|\y)=h(x')[1+\epsilon g(x'|\y)],
\label{conditional_jump}
\end{equation}
where $\epsilon$ is an arbitrary parameter and $h(x')$ is the unconditional jump density ---cf. Eq.~(\ref{h})--- that is related to $h(x'|\y)$ by the constraint
\begin{equation}
h(x')=\int_{-\infty}^{\infty}h(x'|\y)h(\y)d\y.
\label{unconditional}
\end{equation}
Note that this constraint impedes us to consider arbitrary functional forms for $h(x'|\y)$ and $h(x')$. For instance, if we set $h(x'|\y)=[\delta(x'-\lambda \y)+\delta(x'+\lambda \y)]/2$, $\lambda \neq 1$, as in the case of the random walker with shrinking step sizes~\cite{KR04}, Eq.~(\ref{unconditional}) implies $h(x')=\delta(x')$. The distinctive point here is that, unlike geometric random walks, our (unconditional) pdf of $\Delta X_n$ does not depend on $n$. 

Function $g(x'|\y)$ represents the correlation between previous and current jumps and $\epsilon$ governs its strength. Note that in order to meet Eq. (\ref{unconditional}) together with normalization,
$$
\int_{-\infty}^{\infty}h(x'|\y)dx'=1, \qquad\mbox{ for all } \y,
$$
the correlation $g(x'|\y)$ must satisfy (see next section for a general discussion on this issue)
\begin{equation*}
\int_{-\infty}^\infty h(x')g(x'|\y)dx'=\int_{-\infty}^\infty g(x'|\y)h(\y)d\y=0.
\end{equation*}

With the form of $h(x'|\y)$ given in Eq. (\ref{conditional_jump}) the integral equation for the propagator, Eq. (\ref{NMrenewal_3b}), reads
\begin{equation}
\hat{p}(x,s|\y)=\delta(x)\hat{\Psi}(s)+
\hat{\psi}(s)\int_{-\infty}^{\infty}h(x')[1+\epsilon g(x'|\y)]\hat{p}(x-x',s|x')dx'.
\label{NMrenewal_3c}
\end{equation}
We will solve this equation for any even jump density 
\begin{equation}
h(x')=h(-x')
\label{even_h}
\end{equation}
and when the correlation function has the following form
\begin{equation}
g(x'|\y)=\frac{x' \y}{|x'||\y|}=\sgn(x')\sgn(\y),
\label{g_sign}
\end{equation}
meaning that the dependence between current and previous jumps is only through their signs. In other words, the correlation depends on whether consecutive jumps are increasing or decreasing but not on their magnitude. From Eq. (\ref{conditional_jump}) we see that in this case, since $h(x'|\y)$ must be positive definite, $-1\leq \epsilon \leq 1$. In fact, this model might be interpreted as the simplest {\it persistent\/} CTRW~\cite{weissllibre}, in which the probability that the process does not change its direction of movement is equal to $(1+\epsilon)/2$, but where  neither jumps nor sojourns are affected by this persistence. This is the kind of memory we adopted in~\cite{montero_lillo} in order to model the observed anti-correlation, with origin in the bid/ask bounce effect: tick-by-tick price changes tend two oscillate back and forth between to values due to the bid/ask spread.       

Let us note that the functional form of $g(x'|\y)$ given in Eq. (\ref{g_sign}) implies that any dependence on $\y$ is only through $\sgn(\y)=\y/|\y|$. Thus $p(x,t|\y)=p(x,t|\sgn(\y))$ which allows us to write
\begin{equation}
p(x,t|\y)=p^{(+)}(x,t)\Theta(\y)+p^{(-)}(x,t)\Theta(-\y),
\label{total_p}
\end{equation}
where $\Theta(\y)$ is the Heaviside step function. The substitution of Eq. (\ref{total_p}) into Eq. (\ref{NMrenewal_3c}) yields for $\hat{p}^{(\pm)}(x,s)$ the following set of coupled integral equations
\begin{eqnarray*}
\hat{p}^{(+)}(x,s)=\delta(x)\hat{\Psi}(s)&+&
(1+\epsilon)\hat{\psi}(s)\int_{0}^{\infty}h(x')\hat{p}^{(+)}(x-x',s)dx'\nonumber\\
&+&(1-\epsilon)\hat{\psi}(s)\int_{-\infty}^{0}h(x')\hat{p}^{(-)}(x-x',s)dx'
\end{eqnarray*}
\begin{eqnarray*}
\hat{p}^{(-)}(x,s)=\delta(x)\hat{\Psi}(s)&+&
(1-\epsilon)\hat{\psi}(s)\int_{0}^{\infty}h(x')\hat{p}^{(+)}(x-x',s)dx'\nonumber\\
&+&(1+\epsilon)\hat{\psi}(s)\int_{-\infty}^{0}h(x')\hat{p}^{(-)}(x-x',s)dx'.
\end{eqnarray*}
Now the Fourier transform with respect to $x$ turns this set into a system of algebraic equations:
\begin{equation}
\tilde{p}^{(+)}(\omega,s)=\hat{\Psi}(s)+\hat{\psi}(s)\left[(1+\epsilon)\tilde{H}(\omega)\tilde{p}^{(+)}(\omega,s)+
(1-\epsilon)\tilde{H}(-\omega)\tilde{p}^{(-)}(\omega,s)\right],
\label{p+}
\end{equation}
\begin{equation}
\tilde{p}^{(-)}(\omega,s)=\hat{\Psi}(s)+\hat{\psi}(s)\left[(1-\epsilon)\tilde{H}(\omega)\tilde{p}^{(+)}(\omega,s)+
(1+\epsilon)\tilde{H}(-\omega)\tilde{p}^{(-)}(\omega,s)\right],
\label{p-}
\end{equation}
where
\begin{equation*}
\tilde{H}(\omega)\equiv\int_0^\infty e^{i\omega x'}h(x')dx'
\end{equation*}
is the ``half'' Fourier transform of $h(x)$. Obviously
\begin{equation}
\tilde{h}(\omega)=\tilde{H}(\omega)+\tilde{H}(-\omega),
\label{sum_H}
\end{equation}
where $\tilde{h}(\omega)$ is the ``complete'' Fourier transform of $h(x)$. We note that in writing Eqs. (\ref{p+})-(\ref{p-}) and Eq. (\ref{sum_H}) we have imposed the symmetry of $h(x)$ assumed in Eq. (\ref{even_h}). 

Solving for Eqs. (\ref{p+})-(\ref{p-}) we have
\begin{equation}
\tilde{p}^{(\pm)}(\omega,s)=\frac{1-2\epsilon\hat{\psi}(s)\tilde{H}(\mp\omega)}
{1-(1+\epsilon)\hat{\psi}(s)\tilde{h}(\omega)+4\epsilon\hat{\psi}^2(s)|\tilde{H}(\omega)|^2}\hat{\Psi}(s),
\label{fourier_laplace_solution}
\end{equation}
where we have used the fact that for real jump densities $h(x)$ the following identity $\tilde{H}(\omega)\tilde{H}(-\omega)=|\tilde{H}(\omega)|^2$ holds. The final solution to the problem is thus given by the combination of Eqs. (\ref{total_p}) and (\ref{fourier_laplace_solution}).  
Therefore, under the assumptions given in Eqs. (\ref{independent_0}) and (\ref{conditional_jump}) and the special form of the correlation given in Eq. (\ref{g_sign}), we have been able to obtain an exact expression for the Fourier-Laplace transform of the propagator valid for any forms of the waiting time density $\psi(\tau)$ and jump density $h(\y)$, provided that the latter is an even function of $\y$ with no bias. 

Another interesting quantity is the unconditional propagator $p(x,t)$ defined by
\begin{equation*}
p(x,t)=\int_{-\infty}^{\infty}p(x,t|\y)h(\y)d\y.
\end{equation*}
In the analyzed case in which $p(x,t|\y)$ can be decomposed as in Eq. (\ref{total_p}) we have
\begin{equation}
p(x,t)=\frac{1}{2}\left[p^{(+)}(x,t)+p^{(-)}(x,t)\right].
\label{uncond_p}
\end{equation}
From Eqs. (\ref{fourier_laplace_solution}) and (\ref{uncond_p}) we get
\begin{equation}
\tilde{p}(\omega,s)=\frac{1-\epsilon\hat{\psi}(s)\tilde{h}(\omega)}
{1-(1+\epsilon)\hat{\psi}(s)\tilde{h}(\omega)+4\epsilon\hat{\psi}^2(s)|\tilde{H}(\omega)|^2}\hat{\Psi}(s).
\label{fl_uncond}
\end{equation}
Note incidentally that when $\epsilon=0$ this expression reduces to
$$
\tilde{p}(\omega,s)=\frac{\hat{\Psi}(s)}
{1-\hat{\psi}(s)\tilde{h}(\omega)},
$$
which agrees with the solution of the independent case discussed in Sect. \ref{sec2} (cf. Eq. (\ref{solution_0})).

Aside from the unconditional pdf $p(x,t)$, which provides maximal information about the evolution of $X(t)$, there is another quantity of considerable practical interest: the (unconditional) variance of $X(t)$. This quantity has the advantage that it does not require the knowledge of the entire jump distribution $h(\y)$. It suffices to know the pdf  $\psi(\tau)$ and the following two moments of $h(\y)$:
\begin{equation*}
\mu_1\equiv\int_{-\infty}^\infty|\y|h(\y)d\y, \qquad {\rm and} \qquad \mu_2\equiv\int_{-\infty}^\infty \y^2h(\y)d\y.
\end{equation*}
Let $\langle X^2(t)\rangle$ be the unconditional second moment of the process:
$$
\langle X^2(t)\rangle=\int_{-\infty}^{\infty}x^2 p(x,t)dx,
$$
and let us denote by $\hat{m}_2(s)$ its Laplace transform
$$
\hat{m}_2(s)\equiv\int_0^\infty e^{-st}\langle X^2(t)\rangle dt.
$$
This can be written in terms of the joint Fourier-Laplace transform of $p(x,t)$ by
\begin{equation}
\hat{m}_2(s)=-\left.\frac{\partial^2\tilde{p}(\omega,s)}{\partial\omega^2}\right|_{\omega=0}.
\label{fl_moment_def}
\end{equation}
Recall that a direct consequence of the unbiased assumption expressed in Eq. (\ref{even_h}) is that the odd moments of $h(\y)$ are equal to zero. This implies that all odd moments of process $X(t)$ vanish as well; in particular this means that the variance of $X(t)$ coincides with its second moment. The combination of Eqs. (\ref{fl_uncond}) and (\ref{fl_moment_def}) leads, after some manipulations, to the relation
\begin{equation}
\hat{m}_2(s)=\mu_2\frac{\hat{\psi}(s)}{s[1-\hat{\psi}(s)]}+
2\epsilon\mu_1^2\frac{\hat{\psi}^2(s)}{s[1-\hat{\psi}(s)][1-\epsilon\hat{\psi}(s)]}.
\label{fl_moment}
\end{equation}

Let us return to the propagator and particularize to the case of Poissonian waiting times for which $\psi(\tau)=\lambda e^{-\lambda \tau}$ and 
\begin{equation}
\hat{\psi}(s)=\frac{\lambda}{\lambda+s}, \qquad \hat{\Psi}(s)=\frac{1}{\lambda+s}.
\label{poissonian}
\end{equation}
Now Eq. (\ref{fl_uncond}) reads
\begin{equation}
\tilde{p}(\omega,s)=\frac{\lambda+s-\lambda\epsilon\tilde{h}(\omega)}
{(\lambda+s)^2-\lambda(1+\epsilon)(\lambda+s)\tilde{h}(\omega)+4\lambda^2\epsilon|\tilde{H}(\omega)|^2},
\label{fl_uncond_poisson}
\end{equation}
whose inverse Laplace transform yields the unconditional characteristic function \cite{roberts}
\begin{equation}
\tilde{p}(\omega,t)=e^{-\lambda t[1-(1+\epsilon)\tilde{h}(\omega)/2]}
\left\{\cosh[\lambda t\tilde{k}(\omega)/2]+
(1-\epsilon)\frac{\tilde{h}(\omega)}{\tilde{k}(\omega)}\sinh[\lambda t\tilde{k}(\omega)/2]\right\},
\label{cf}
\end{equation}
where
\begin{equation}
\tilde{k}(\omega)\equiv\sqrt{(1+\epsilon)^2\tilde{h}^2(\omega)-16\epsilon|\tilde{H}(\omega)|^2}.
\label{k}
\end{equation}

As to the second moment is concern Eq. (\ref{fl_moment}) can be inverted at once with the result
\begin{equation}
\langle X^2(t)\rangle=\mu_2\lambda t+
\frac{2\mu_1^2\epsilon}{(1-\epsilon)^2}\left[\lambda(1-\epsilon)t+e^{-\lambda(1-\epsilon)t}-1\right].
\label{variance}
\end{equation}
Observe that in the independent case ($\epsilon=0$) the variance shows an ordinary diffusion behavior while correlations introduce a richer dynamics. 

In order to invert Eq. (\ref{cf}) and thus obtaining an expression for the propagator $p(x,t)$ we must chose a functional form for the jump density $h(\y)$. We will select the two-sided exponential density,
\begin{equation*}
h(\y)=(\gamma/2)e^{-\gamma|\y|},
\end{equation*}
for two main reasons. On the one hand it can be of interest in finance, the field that motivates this work in the first instance. Even though it is well established that pdf's of financial returns show a power-law decay~\cite{GPAMS99}, there are an increasing number of evidences pointing to the fact that small and moderate returns are better described through a Laplace law ---see~\cite{SY07} and references therein. On the other hand, one of the main motivations of this sections is the introduction and subsequent analysis of a case for which we can obtain closed expressions. It is clear from Eq.~(\ref{cf}) that when the characteristic function of $h(\y)$ is intricate this goal will be well out of reach. In our case $\tilde{h}(\omega)$ is the inverse of a polynomial:  
\begin{equation}
\tilde{h}(\omega)=\frac{\gamma^2}{\gamma^2+\omega^2}, \qquad \tilde{H}(\omega)=\frac{\gamma/2}{\gamma-i\omega}.
\label{f_laplacian}
\end{equation}

We incidentally note that now the variance of the process is given by Eq. (\ref{variance}) where $\mu_1=1/\gamma$ and $\mu_2=2/\gamma^2$. When $0<\epsilon\leq 1$ one can show that the Fourier inversion of Eq. (\ref{cf}) reads~\footnote{In order to get Eqs. (\ref{explicit})-(\ref{explicit_0}), as well as all the numerical results presented below, it is somewhat simpler to start from Eq. (\ref{fl_uncond_poisson}) and use Eq. (\ref{f_laplacian}) to invert first the Fourier transform and finally the Laplace transform.}
\begin{eqnarray}
p(x,t)&=&e^{-\lambda t}\delta(x)+\frac{\gamma e^{-\lambda t}}{2\sqrt{\epsilon}}
\int_{\lambda\gamma|x|\sqrt{\epsilon}}^\infty  \left(\frac{t}{u}\right)^{1/2}e^{-(1+\epsilon)u/2\lambda\epsilon}\Biggl[I_1(2\sqrt{u t})\nonumber\\
&-&
\lambda\epsilon\left(\frac{t}{u}\right)^{1/2}I_2(2\sqrt{u t})\Biggr]I_0\left[\frac{(1-\epsilon)}{2\lambda\epsilon}\sqrt{u^2-\epsilon\lambda^2\gamma^2|x|^2}\right]du,
\label{explicit}
\end{eqnarray}
($0<\epsilon\leq 1$), where $I_n(z)$ are modified Bessel functions. Although the case $\epsilon=1$ is contained in Eq. (\ref{explicit}) it can be written more explicitly as
\begin{equation}
p(x,t)=e^{-\lambda t}\delta(x)+\frac{\gamma}{2}\sqrt{\frac{\lambda t}{\gamma|x|}}I_1\left(2\sqrt{\gamma|x|\lambda t}\right) e^{-\lambda t-\gamma|x|}.
\label{explicit_1}
\end{equation}
The recovery of the independent case $\epsilon=0$ from Eq.~(\ref{explicit}) is a delicate issue that deserves a special treatment. In this case one has
\begin{equation}
p(x,t)=e^{-\lambda t}\delta(x)+\gamma\left(\frac{\lambda t}{\pi}\right)^{1/2}
\int_{0}^\infty \frac{I_1(u)}{u}\exp\left\{-\frac{u^2}{4\lambda t}-\frac{\gamma^2|x|^2\lambda t}{u^2}-\lambda t\right\}du,
\label{explicit_0}
\end{equation}
($\epsilon=0$).

\begin{figure}[tbhp] 
\begin{tabular}{rlrl}
(a)&{\hfil \includegraphics[width=0.45\textwidth,keepaspectratio=true]{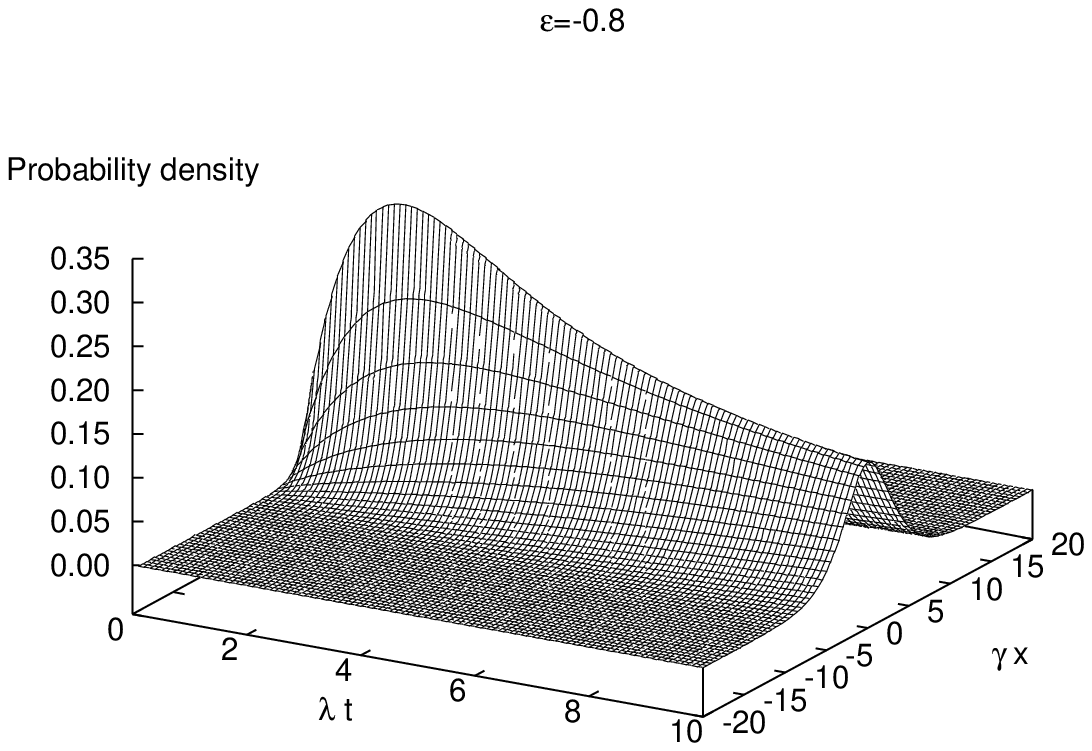}}&(b)&{\hfill \includegraphics[width=0.45\textwidth,keepaspectratio=true]{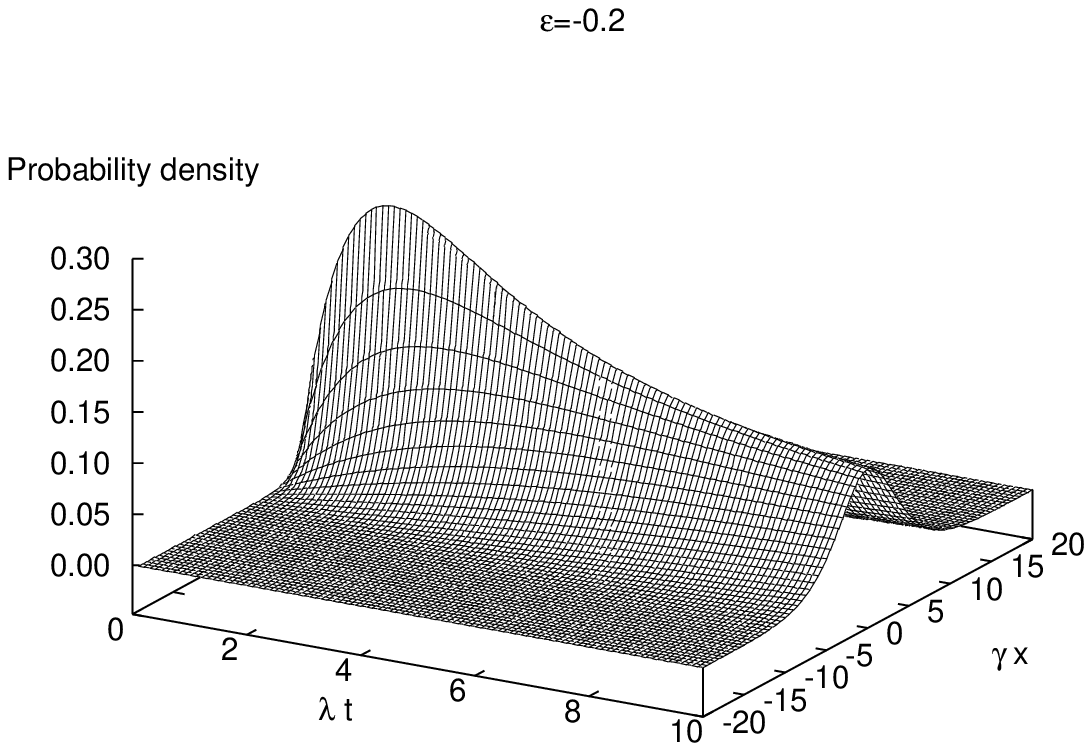}}\\ 
(c)&{\hfil \includegraphics[width=0.45\textwidth,keepaspectratio=true]{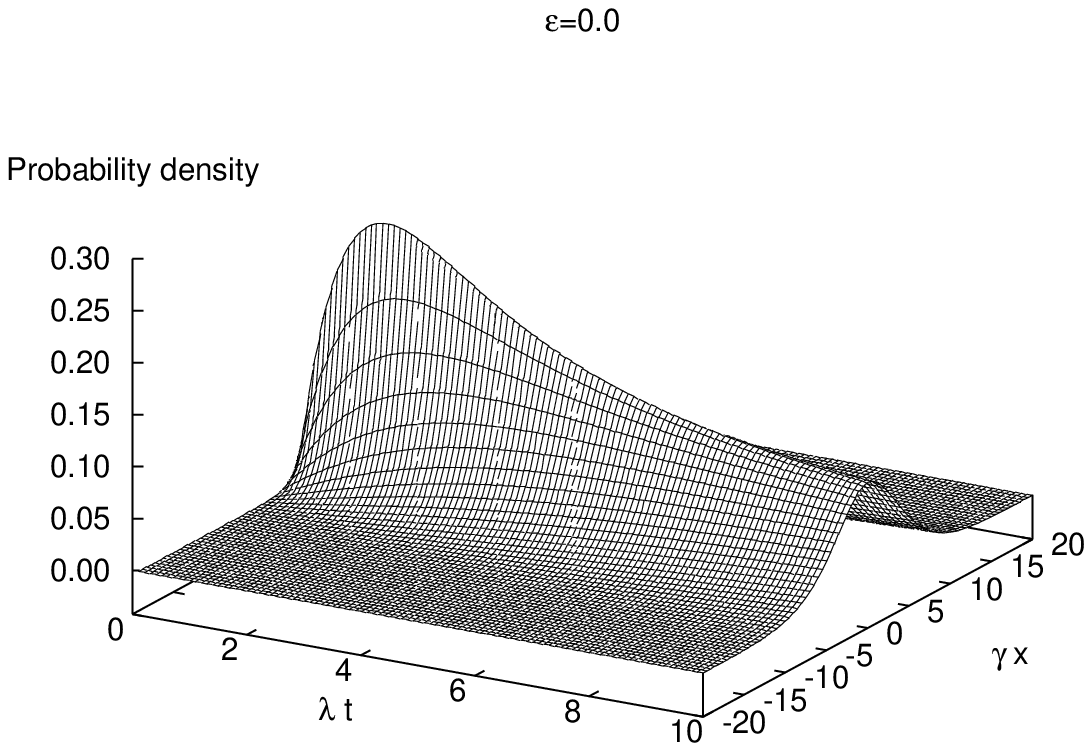}}&(d)&{\hfill \includegraphics[width=0.45\textwidth,keepaspectratio=true]{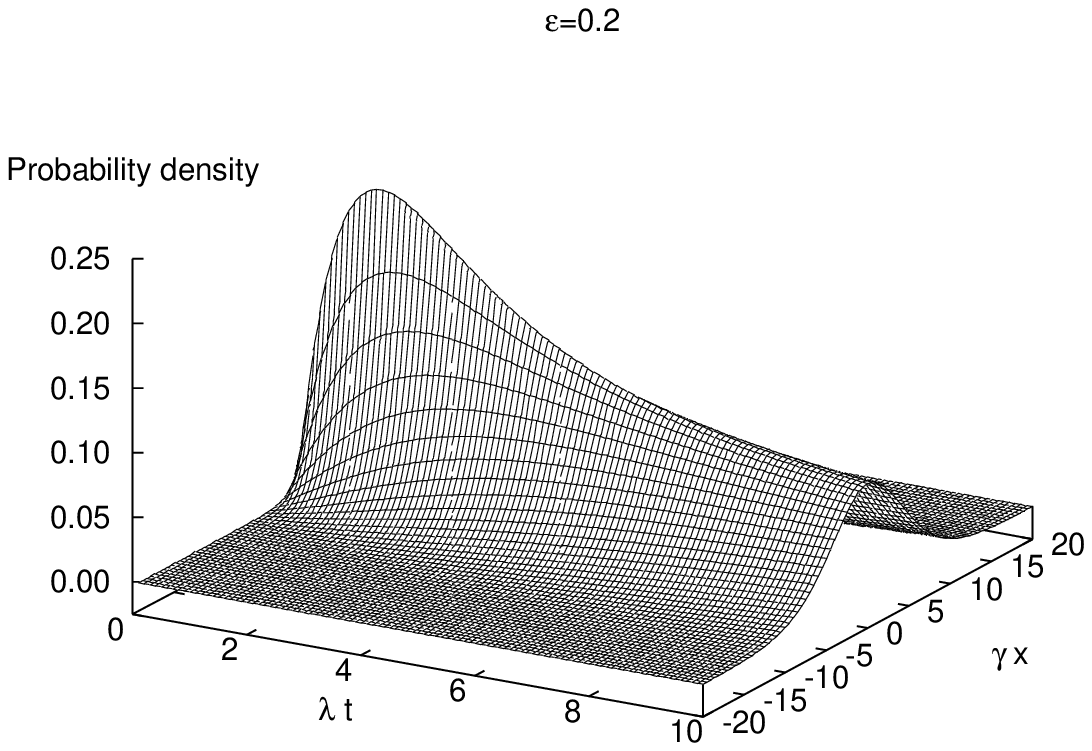}}\\ 
(e)&{\hfil \includegraphics[width=0.45\textwidth,keepaspectratio=true]{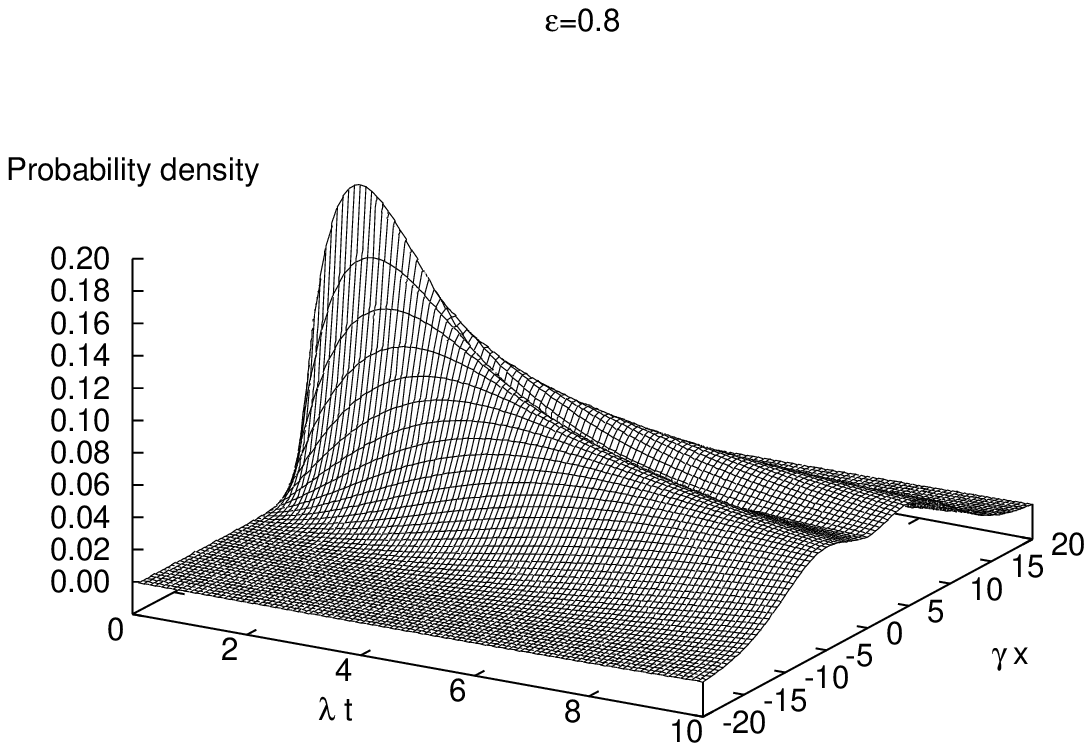}}&(f)&{\hfill \includegraphics[width=0.45\textwidth,keepaspectratio=true]{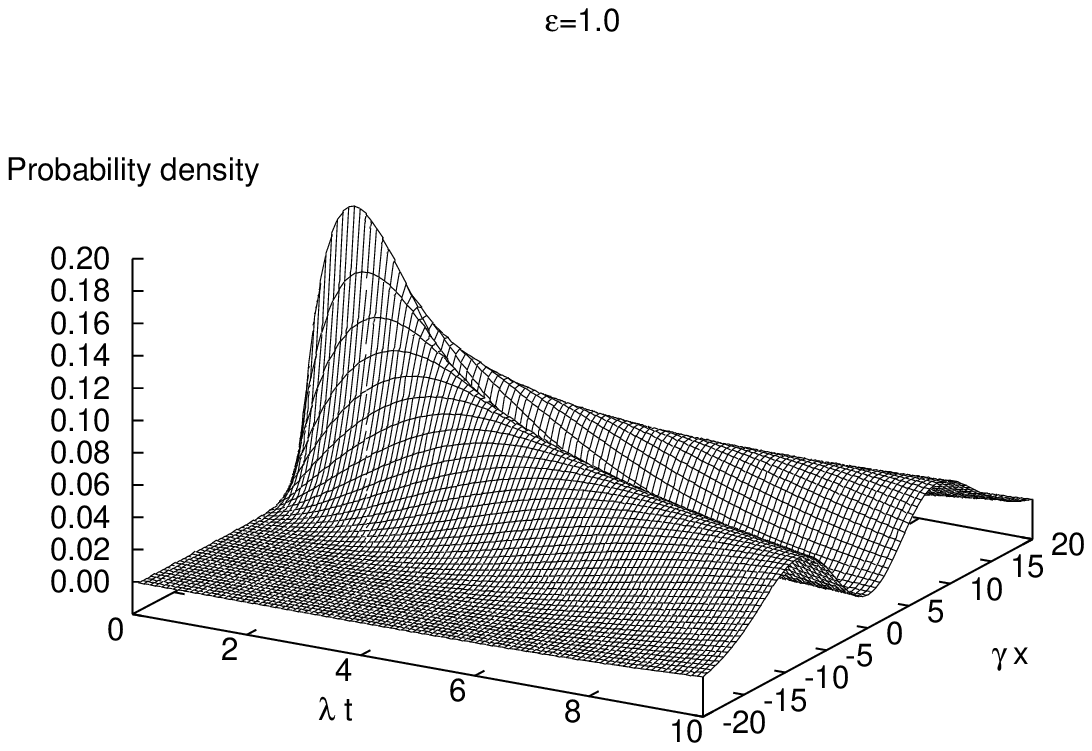}}\\ 
\end{tabular}
\caption{Probability density function for different degrees of correlation. We can see how the negative correlated processes concentrate the probability around the origin, whereas the positive correlated processes spread faster. Within this context, if the correlation is large enough, the system shows a transient bimodality.}
\label{Fig1}
\end{figure}

We can now graphically explore some of the most relevant properties of the unconditional propagator in the analyzed example. However, as we will show, our example will share those traits with any process that presents Poissonian waiting times, {\it i.e.\/} for which Eq.~(\ref{variance}) stands. In Fig.~\ref{Fig1} we plot the regular part of the probability density function (in $\gamma$ units) for different values of $\epsilon$. In particular we present the cases of (i) a strong anti-correlated process ($\epsilon$ close to $-1$), Fig.~\ref{Fig1}.a; (ii) a weak anti-correlated process, Fig.~\ref{Fig1}.b; (iii) the independent case ($\epsilon=0$), Fig.~\ref{Fig1}.c; (iv) a weak correlated process, Fig.~\ref{Fig1}.d; (v) a strong correlated process ($\epsilon$ close to $1$), Fig.~\ref{Fig1}.e; and finally, (vi) the completely persistent case ($\epsilon=1$), Fig.~\ref{Fig1}.f. The visible effect of anti-persistent memory in the process is that the probability density function becomes narrower around $x=0$. Thus, for larger negative values of $\epsilon$, the system tends to remain longer near the origin, and the process exhibits sub-diffusive behavior for small timescales. In fact, from Eq.~(\ref{variance}) we will have for $\lambda t \ll 1$ that
\begin{equation*}
\sigma(t)=\sqrt{\langle X^2(t)\rangle} \approx \sqrt{\mu_2 \lambda t + \epsilon \mu^2_1 \lambda^2 t^2},
\end{equation*}
and therefore the process is sub-diffusive for $\epsilon<0$, diffusive in the independent case (recall that this statement is valid for all timescales), and super-diffusive when $\epsilon>0$. 

\begin{figure}[tbhp] 
\begin{tabular}{rlrl}
(a)&{\hfil \includegraphics[width=0.45\textwidth,keepaspectratio=true]{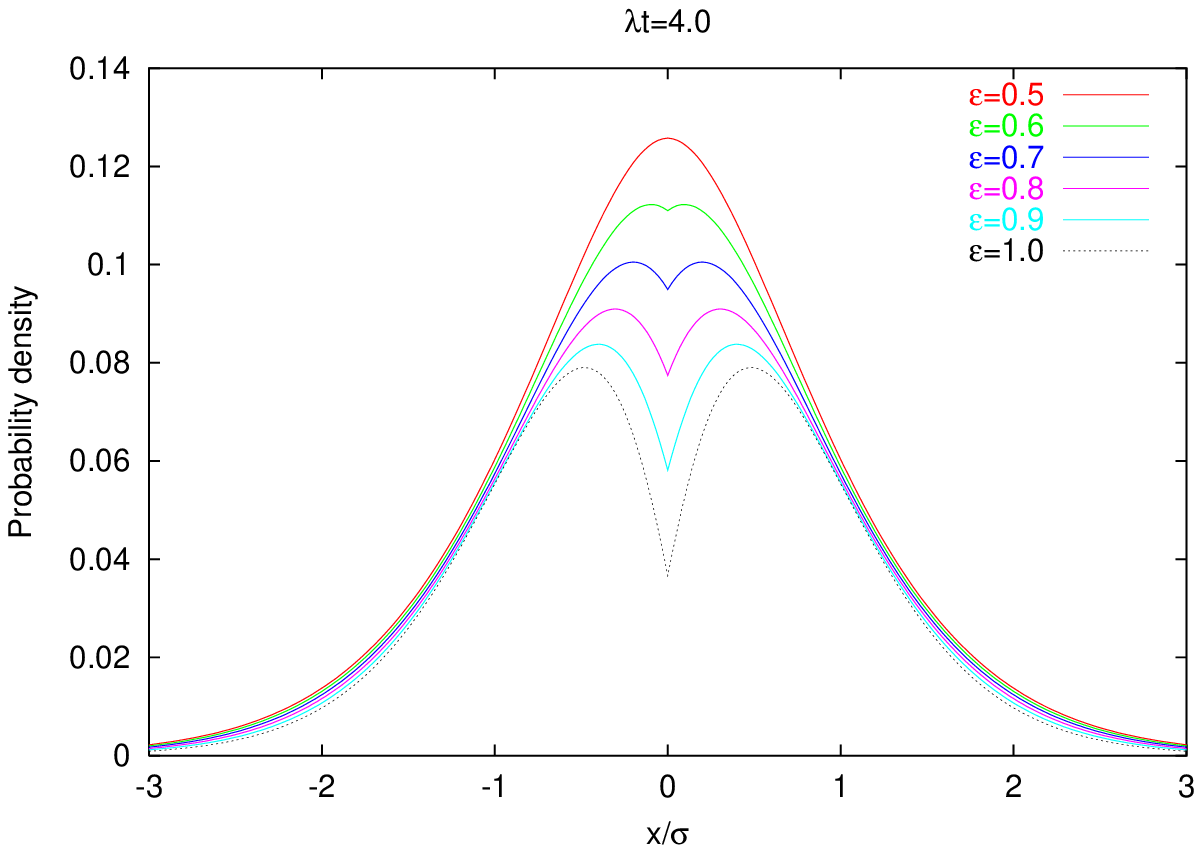}}&(b)&{\hfill \includegraphics[width=0.45\textwidth,keepaspectratio=true]{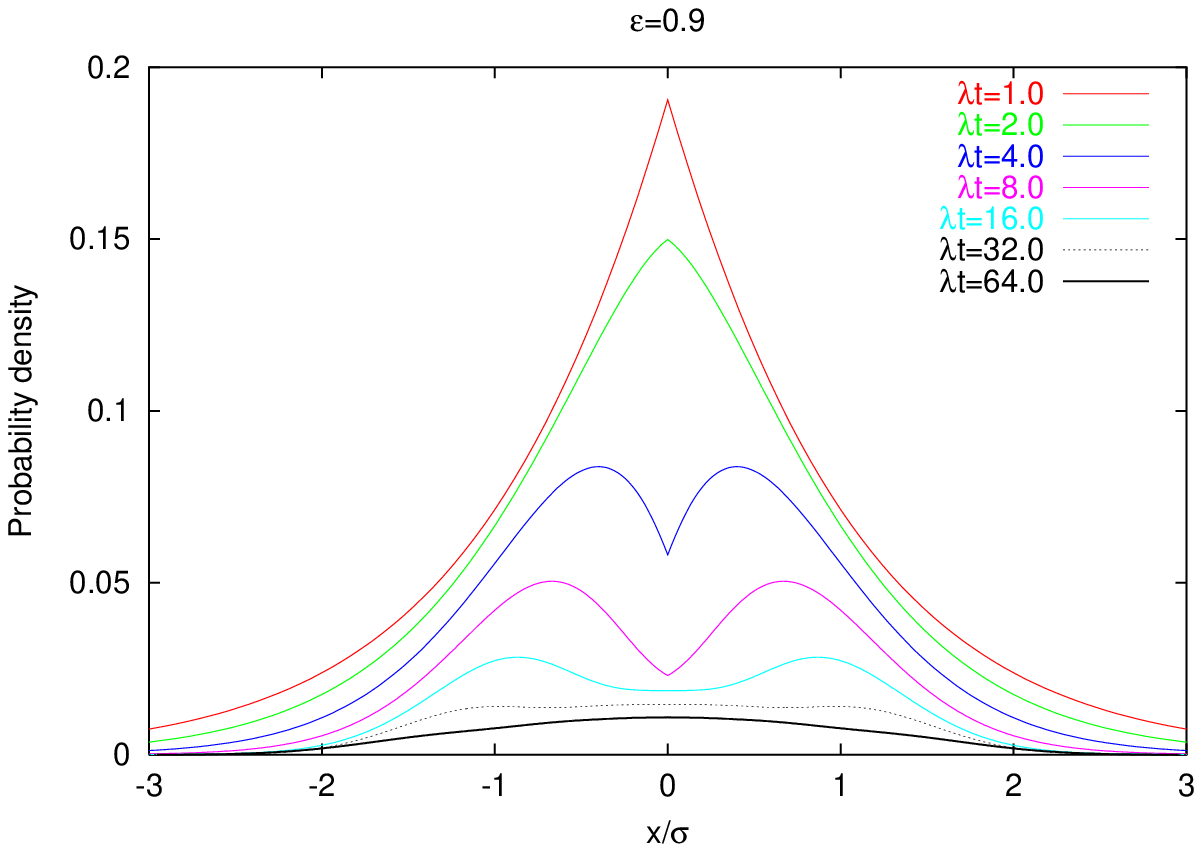}}
\end{tabular}
\caption{(Color online) The role of $\epsilon$ and $t$ in the bimodality of the probability density function. In (a) we can see how bimodality becomes more evident for larger values of $\epsilon$. In (b) we can check how this feature finally disappears even in a strongly correlated process. We have used the standard deviation of the process, $\sigma(t)$, in order to make the plots commensurable.}
\label{Fig2}
\end{figure}

For large positive values of the correlation parameter $\epsilon$, see Fig.~\ref{Fig1}.e and Fig.~\ref{Fig1}.f, the unconditional propagator presents another interesting property: there are two modes in the probability density function. These two modes are located in the vicinity of the conditional first moments $\langle X^{(\pm)}(t)\rangle$,  
$$
\langle X^{(\pm)}(t)\rangle=\int_{-\infty}^{\infty}x p^{(\pm)}(x,t)dx=\pm \mu_1 \frac{\epsilon}{1-\epsilon} \left[ 1- e^{-\lambda(1-\epsilon)t}\right],
$$
and become more and more notorious for increasing values of $\epsilon$, as we show in Fig.~\ref{Fig2}.a. However, this apparent bimodality must disappear for large timescales, since if $\lambda t \gg 1$ we have~\footnote{Note that the expression inside the square root is positive definite even when $-1\leq \epsilon <0$ because the Cauchy-Schwarz inequality implies that $\mu_1^2<\mu_2$ for $h(x')\neq[\delta(x'-c)+\delta(x+'c)]/2$. The latter case corresponds to a process that moves back and forth between two fixed points, when $\epsilon=-1$.}
\begin{equation*}
\sigma(t) \sim \sqrt{\left[\mu_2+ \frac{2\mu_1^2\epsilon}{1-\epsilon}\right]\lambda t},
\end{equation*}
whenever $\epsilon \neq 1$. This means that we must eventually attain to diffusive limit, as it is depicted in Fig.~\ref{Fig2}.b. Only in the completely persistent case ($\epsilon=1$) the phenomenon is not of transient nature, because in this case the process is super-diffusive at all timescales:
$$
\sigma(t) = \sqrt{\mu_2\lambda t+ \mu_1^2 \lambda^2 t^2}.
$$ 

\section{A weakly dependent model}
\label{sec4}

Obtaining exact expressions for the propagator is usually quite involved, not to say impossible, in many practical situations and we have to resort to approximations based on perturbation solutions of the integral equation (\ref{NMrenewal}). With this purpose in mind we will work with the following  form of the joint density that generalizes the jump density given in Eq. (\ref{conditional_jump}):
\begin{equation}
\rho(x',t'|\y,\tau)=\rho(x',t')[1+\epsilon g(x',t'|\y,\tau)],
\label{wrho}
\end{equation}
 where $\rho(x,t)$ is the unconditional joint density satisfying
\begin{equation}
\rho(x',t')=\int_0^\infty d\tau\int_{-\infty}^{\infty} \rho(x',t'|\y,\tau)\rho(\y,\tau)d\y,
\label{un_rho}
\end{equation}
and the function $g(x',t'|\y,\tau)$ indicates the correlations between the waiting time and the jump of the current sojourn and those of the preceding sojourn. $\epsilon$ is a parameter measuring the strength of this correlation. In what follows we will suppose that $\epsilon$ is small, {\it i.e.\/}, the model is weakly dependent. Function $g$ is not arbitrary and must satisfy two consistency conditions. Indeed, from the normalization of the densities $\rho(x',t'|\y,\tau)$ and $\rho(x',t')$:
$$
\int_0^\infty dt'\int_{-\infty}^{\infty}\rho(x',t'|\y,\tau)dx'=\int_0^\infty dt'\int_{-\infty}^{\infty}\rho(x',t')dx'=1,
$$
immediately follows that
\begin{equation}
\int_0^\infty dt'\int_{-\infty}^{\infty}\rho(x',t') g(x',t'|\y,\tau)dx'=0.
\label{consistency_1}
\end{equation}
On the other hand, plugging Eq. (\ref{wrho}) into Eq. (\ref{un_rho}) and taking again into account normalization we get
\begin{equation}
\int_0^\infty d\tau\int_{-\infty}^{\infty}\rho(\y,\tau) g(x',t'|\y,\tau)d\y=0.
\label{consistency_2}
\end{equation}
We also observe that if 
$\rho(x',t') g(x',t'|\y,\tau)$ and $\rho(\y,\tau) g(x',t'|\y,\tau)$ are integrable functions with respect to $t'$ and $\tau$ respectively, then for consistency conditions, Eqs. (\ref{consistency_1})-(\ref{consistency_2}), to hold it suffices that:
\begin{equation}
\int_{-\infty}^{\infty}\rho(x',t') g(x',t'|\y,\tau)dx=\int_{-\infty}^{\infty} g(x',t'|\y,\tau)\rho(\y,\tau)d\y=0.
\label{consistency}
\end{equation}

The starting point of our analysis is the renewal equation for the joint Fourier-Laplace transform of the propagator $\tilde{p}(\omega,s|\y,\tau)$. Substituting then Eq. (\ref{wrho}) into Eq. (\ref{NMrenewal_2}) we have
\begin{eqnarray}
\tilde{p}(\omega,s|\y,\tau)&=&\tilde{p}_0(\omega,s|\y,\tau) \nonumber \\
&+&\int_0^\infty dt' e^{-st'} \int_{-\infty}^{\infty} e^{i\omega x'}\rho(x',t')[1+\epsilon g(x',t'|\y,\tau)] \tilde{p}(\omega,s|x',t')dx'.
\label{w_eq_1}
\end{eqnarray}
Assuming that $\epsilon$ is small we look for a solution to this equation in the form
\begin{equation}
\tilde{p}(\omega,s|\y,\tau)=\tilde{q}(\omega,s)+\sum_{n=1}^\infty \epsilon^n \tilde{p}^{(n)}(\omega,s|\y,\tau).
\label{series}
\end{equation}

In order to proceed further we need to know the dependence on $\epsilon$ of the propagator prior the first sojourn. From Eqs. (\ref{NMp0})-(\ref{NMPsi}) and Eq. (\ref{wrho}) we write
$$
p_0(x,t|\y,\tau)=\delta(x)\int_t^\infty dt'\int_{-\infty}^{\infty}\rho(x',t')[1+\epsilon g(x',t'|\y,\tau)]dx',
$$
but
$$
\int_t^\infty dt'\int_{-\infty}^{\infty}\rho(x',t')dx'=\Psi(t)
$$
and using the consistency condition (\ref{consistency}) we finally get
\begin{equation}
p_0(x,t|\y,\tau)=\delta(x)\Psi(t),
\label{wp0}
\end{equation}
hence $p_0$ is independent of $\epsilon$. 

One can easily see by substituting Eqs. (\ref{series}) and (\ref{wp0}) into Eq. (\ref{w_eq_1}) that the zeroth-order $\tilde{q}(\omega,s)$ corresponds to the independent case 
\begin{equation*}
\tilde{q}(\omega,s)=\frac{\hat{\Psi}(s)}{1-\tilde{\rho}(\omega,s)},
\end{equation*}
while for $n=1,2,3,\cdots$ we have the recursive integral equations:
\begin{eqnarray}
\tilde{p}^{(n)}(\omega,s|\y,\tau)&=&Q^{(n-1)}(\omega,s|\y,\tau)\nonumber \\
&+&\int_0^\infty dt' e^{-st'} \int_{-\infty}^{\infty} e^{i\omega x'}\rho(x',t')\tilde{p}^{(n)}(\omega,s|x',t')dx',
\label{int_eq_n}
\end{eqnarray}
where
\begin{equation}
Q^{(n-1)}(\omega,s|\y,\tau)= \int_0^\infty dt' e^{-st'} \int_{-\infty}^{\infty} e^{i\omega x'}\rho(x',t')g(x',t'|\y,\tau)
\tilde{p}^{(n-1)}(\omega,s|x',t')dx',
\label{Q^n}
\end{equation}
and
\begin{equation}
Q^{(0)}(\omega,s|\y,\tau)=\tilde{q}(\omega,s)\int_0^{\infty}dt' e^{-st'}\int_{-\infty}^{\infty}e^{i\omega x'}\rho(x',t')g(x',t'|\y,\tau)dx'
\label{Q^0}
\end{equation}

In the Appendix \ref{Appendix_B} we show that the solution to Eq. (\ref{int_eq_n}) is given by the recursive expression ($n=1,2,3,\cdots$)
\begin{eqnarray}
\tilde{p}^{(n)}(\omega,s|\y,\tau)&=&\int_0^\infty dt' e^{-st'} \int_{-\infty}^{\infty} e^{i\omega x'}\rho(x',t')\biggl[g(x',t'|\y,\tau)
\nonumber\\
&+&\frac{1}{1-\tilde{\rho}(\omega,s)}G(x',t'|\omega,s)\biggr]\tilde{p}^{(n-1)}(\omega,s|x',t')dx',
\label{recursive_n}
\end{eqnarray}
where
\begin{equation}
G(x',t'|\omega,s)\equiv \int_0^\infty d\tau e^{-s\tau}\int_{-\infty}^{\infty}e^{i\omega y}\rho(y,\tau)g(x',t'|\y,\tau)d\y,
\label{G}
\end{equation}
and
$$
\tilde{p}^{(0)}(\omega,s|x',t')=\tilde{q}(\omega,s)=\frac{\hat{\Psi}(s)}{1-\tilde{\rho}(\omega,s)}.
$$

For $n=1$ we have
\begin{eqnarray*}
\tilde{p}^{(1)}(\omega,s|\y,\tau)&=&\tilde{q}(\omega,s)\int_0^\infty dt' e^{-st'} \int_{-\infty}^{\infty} e^{i\omega x'}\rho(x',t')\biggl[g(x',t'|\y,\tau)
\\
&+&\frac{1}{1-\tilde{\rho}(\omega,s)}G(x',t'|\omega,s)\biggr]dx',
\end{eqnarray*}
which, after defining
\begin{equation}
\tilde{g}_\rho(\omega,s|\y,\tau)\equiv\int_0^\infty dt' e^{-st'} \int_{-\infty}^{\infty} e^{i\omega x'}\rho(x',t')g(x',t'|\y,\tau)dx'
\label{tilde_g}
\end{equation}
and (cf. Eq. (\ref{G}))
\begin{equation}
\tilde{G}_\rho(\omega,s)\equiv\int_0^\infty dt' e^{-st'} \int_{-\infty}^{\infty} e^{i\omega x'}\rho(x',t')G(x',t'|\omega,s)dx',
\label{tilde_G}
\end{equation}
can be written as
$$
\tilde{p}^{(1)}(\omega,s|\y,\tau)=\biggl[\tilde{g}_\rho(\omega,s|\y,\tau)+
\frac{1}{1-\tilde{\rho}(\omega,s)}\tilde{G}_\rho(\omega,s)\biggr]\tilde{q}(\omega,s). 
$$
Therefore, the joint Fourier-Laplace transform of the propagator up to first order in $\epsilon$ is
\begin{equation}
\tilde{p}(\omega,s|\y,\tau)=\Biggl\{1+\epsilon\biggl[\tilde{g}_\rho(\omega,s|\y,\tau)
+
\frac{\tilde{G}_\rho(\omega,s)}{1-\tilde{\rho}(\omega,s)}\biggr]+{\rm O}(\epsilon^2)\Biggr\}\frac{\hat{\Psi}(s)}{1-\tilde{\rho}(\omega,s)}.
\label{first_order}
\end{equation}

From the above expression of the conditional propagator we can also get the unconditional propagator defined by
\begin{equation}
p(x,t)=\int_0^\infty d\tau \int_{-\infty}^{\infty} \rho(\y,\tau)p(x,t|\y,\tau)d\y.
\label{uncond_def}
\end{equation}
From Eq. (\ref{first_order}) and taking into account the normalization of $\rho(\y,\tau)$ we have
$$
\tilde{p}(\omega,s)=\Biggl\{1+\epsilon\Biggl[\int_0^\infty d\tau \int_{-\infty}^{\infty} \rho(\y,\tau)\tilde{g}_\rho(\omega,s|\y,\tau)
+
\frac{\tilde{G}_\rho(\omega,s)}{1-\tilde{\rho}(\omega,s)}\Biggr]+{\rm O}(\epsilon^2)\Biggr\}\frac{\hat{\Psi}(s)}{1-\tilde{\rho}(\omega,s)}.
$$
But from Eqs. (\ref{consistency_2}) and (\ref{tilde_g}) one can easily see that
$$
\int_0^\infty d\tau \int_{-\infty}^{\infty} \rho(\y,\tau)\tilde{g}_\rho(\omega,s|\y,\tau)d\y=0.
$$
Hence
\begin{equation}
\tilde{p}(\omega,s)=\left[1+\epsilon
\frac{\tilde{G}_\rho(\omega,s)}{1-\tilde{\rho}(\omega,s)}+
{\rm O}(\epsilon^2)\right]\frac{\hat{\Psi}(s)}{1-\tilde{\rho}(\omega,s)}.
\label{uncond_1}
\end{equation}

We will finally present an instrumental example of the use of the perturbation technique just developed. We assume, as in Sect. \ref{sec3b}, that the joint density $\rho$ factorizes as (cf. Eqs. (\ref{independent_0}) and (\ref{conditional_jump}))
\begin{equation}
\rho(x',t'|\y,\tau)=\psi(t')h(x')[1+\epsilon g(x'|\y)],
\label{independent_0b}
\end{equation}
where $|\epsilon|\ll 1$ and $h(\y)$ is an even function of $\y$. Contrary to the solvable case discussed in Sect. \ref{sec3b} in which the correlation $g(x'|\y)$ depends solely on the signs of consecutive jumps (cf. Eq. (\ref{g_sign})), we now assume that the correlation depends also on jump sizes. We thus suppose
\begin{equation}
g(x'|\y)=\sgn(x')e^{-a(|x'|+|\y|)}\sgn(\y), 
\label{corr_pet}
\end{equation}
($a\geq 0$). With this correlation we will evaluate the expressions for the propagators $p(x,t|\y)$ and $p(x,t)$ as given respectively by Eqs. (\ref{first_order}) and (\ref{uncond_1}) for their joint Fourier-Laplace transform up to first order in $\epsilon$. To this end we need the explicit expressions for the auxiliary quantities $\tilde{g}_\rho(\omega,s|\y)$ and $\tilde{G}_\rho(\omega,s)$ which appear in Eqs. (\ref{first_order}) and (\ref{uncond_1}). 

Using Eqs. (\ref{independent_0b}) and (\ref{corr_pet}), the expression for $\tilde{g}_\rho(\omega,s|\y)$ defined in Eq. (\ref{tilde_g}) can be written as 
$$
\tilde{g}_\rho(\omega,s|\y)=\hat{\psi}(s)\sgn(\y)e^{-a|\y|}\int_{-\infty}^{\infty}\sgn(x')e^{i\omega x'-a|x'|}h(x')dx'.
$$
But taking into account the symmetry of $h(x')$ expressed by Eq. (\ref{even_h}) we can write
$$
\int_{-\infty}^{\infty}\sgn(x') e^{i\omega x'-a|x'|} h(x')dx'=2i \int_{0}^{\infty}e^{-ax'}h(x')\sin\omega x' dx'.
$$
Hence 
\begin{equation}
\tilde{g}_\rho(\omega,s|\y)=2i\hat{\psi}(s)\tilde{h}_{\rm s}(\omega,a)e^{-a|\y|}\sgn(\y),
\label{tilde_g(2)}
\end{equation}
where
\begin{equation}
\tilde{h}_{\rm s}(\omega,a)\equiv\int_{0}^{\infty}e^{-ax'}h(x')\sin\omega x' dx'
\label{H_s}
\end{equation}
is the Fourier sine transform of $e^{-ax'}h(x')$. 

Proceeding in an analogous way we see that the expression for $\tilde{G}_\rho(\omega,s)$ defined in Eq. (\ref{tilde_G}) is given by
$$
\tilde{G}_\rho(\omega,s)=\tilde{\psi}^2(s)\left[\int_{-\infty}^\infty \sgn(x')h(x')e^{i\omega x'-a|x'|}dx'\right]^2.
$$
But as we have just seen (cf. Eq. (\ref{H_s}))  
$$
\int_{-\infty}^\infty \sgn(x')h(x')e^{i\omega x'-a|x'|}dx=2i\tilde{h}_{\rm s}(\omega,a);
$$
whence 
\begin{equation}
\tilde{G}_\rho(\omega,s)=-4\tilde{\psi}^2(s)\tilde{H}_{\rm s}^2(\omega,s).
\label{tilde_G(2)}
\end{equation}
Substituting Eqs. (\ref{tilde_g(2)}) and (\ref{tilde_G(2)}) into Eq. (\ref{first_order}) yields
\begin{eqnarray*}
\tilde{p}(\omega,s|\y)=\Biggl\{1&+&\epsilon\tilde{\psi}(s)\tilde{h}_{\rm s}(\omega,a)\biggl[2ie^{-a|\y|}\sgn(\y)
\nonumber \\
&-&
\frac{4\tilde{\psi}(s)\tilde{h}_{\rm s}(\omega,a)}{1-\tilde{\psi}(s)\tilde{h}(\omega)}\biggr]+
{\rm O}(\epsilon^2)\Biggr\}\frac{\hat{\Psi}(s)}{1-\tilde{\psi}(s)\tilde{h}(\omega)}.
\end{eqnarray*}

For Poissonian sojourns (cf. Eq. (\ref{poissonian})) we can take the inverse Laplace transform of this expression which yields the conditional characteristic function:
\begin{eqnarray}
\tilde{p}(\omega,t|\y)&=&e^{-\lambda t[1-\tilde{h}(\omega)]}
+\epsilon\Biggl\{2i\frac{\tilde{h}_{\rm s}(\omega,a)}{\tilde{h}(\omega)}\Bigl[-e^{-\lambda t}+
e^{-\lambda t[1-\tilde{h}(\omega)]}\Bigr]e^{-a|\y|}\sgn(\y) \nonumber \\
&-&4\biggl[\frac{\tilde{h}_{\rm s}(\omega,a)}{\tilde{h}(\omega)}\biggr]^2
\Bigl[e^{-\lambda t}+[\lambda t \tilde{h}(\omega)-1]e^{-\lambda t[1-\tilde{h}(\omega)]}\Bigr]\Biggr\}
+{\rm O}(\epsilon^2).
\label{first_order(3)}
\end{eqnarray}
The unconditional characteristic function can be analogously obtained through Eqs. (\ref{uncond_1}) and (\ref{tilde_G(2)}) or else directly by substituting Eq. (\ref{first_order(3)}) into
$$
\tilde{p}(\omega,t)=\int_{-\infty}^{\infty}\tilde{p}(\omega,t|\y)h(\y)d\y;
$$
by either way one chooses the final result is
\begin{equation*}
\tilde{p}(\omega,t)=e^{-\lambda t[1-\tilde{h}(\omega)]}
-4\epsilon\biggl[\frac{\tilde{h}_{\rm s}(\omega,a)}{\tilde{h}(\omega)}\biggr]^2
\Bigl[e^{-\lambda t}+[\lambda t \tilde{h}(\omega)-1]e^{-\lambda t[1-\tilde{h}(\omega)]}\Bigr]
+{\rm O}(\epsilon^2).
\end{equation*}

For this example the variance of the process, 
$\langle X^2(t)\rangle=-\partial^2\tilde{p}(\omega,s)/\partial\omega^2|_{\omega=0}$, is easily seen to be
\begin{equation*}
\langle X^2(t)\rangle=\lambda t\mu_2+8\epsilon\kappa_a^2(e^{-\lambda t}+\lambda t-1),
\end{equation*}
where
$$
\mu_2=-\tilde{h}''(0)\qquad{\rm and}\qquad \kappa_a=\tilde{h}_{\rm s}'(0,a).
$$

\section{Summary and conclusions}
\label{sec5}

We have presented a generalization of the CTRW which include correlations between consecutive sojourns and jumps. We have derived the general equations governing the time evolution of the dependent walk and we have exactly solved them in some particular instances. We have also developed a general perturbation technique aimed to treat, within any desired degree of accuracy, weakly dependent models. That is, those models in which there is a low correlation between consecutive events. 

Due to the extensive analytical apparatus and technical aspects contained in this paper, which may obscure the main objective and perhaps discourage potential users of the technique presented, we shall now summarize the key expressions of our development.

The model is based on a two-dimensional Markov series of jumps and sojourns, with a conditional joint density $\rho(\xi',\tau'|\xi,\tau)$ defined in Eq.~(\ref{rho_mark}). 
The main objective of CTRW is obtaining the so-called propagator $p(x,t|\y,\tau)$, that is, the (conditional) probability density function of the process $X(t)$, provided we know the value of the last jump size, $\xi$, and waiting time, $\tau$. The propagator obeys a renewal equation (cf. Eq. (\ref{NMrenewal})) and its Fourier-Laplace transform satisfies the integral equation~(\ref{NMrenewal_2}). We have been able to find an exact solution to this equation when the joint density has the following form (cf. Eqs. (\ref{independent_0}) and (\ref{conditional_jump}))
\begin{equation}
\rho(\xi',\tau'|\y,\tau)=\psi(\tau')h(\xi')[1+\epsilon\ \sgn(\xi')\sgn(\y)],
\label{c2}
\end{equation}
for which the correlation between jumps depend on whether they are increasing or decreasing but not on their magnitude. In such a case the Fourier-Laplace transform of the unconditional propagator
is given by (cf. Eq. (\ref{fl_uncond}))
$$
\tilde{p}(\omega,s)=\frac{1-\epsilon\hat{\psi}(s)\tilde{h}(\omega)}
{1-(1+\epsilon)\hat{\psi}(s)\tilde{h}(\omega)+4\epsilon\hat{\psi}^2(s)|\tilde{H}(\omega)|^2}\hat{\Psi}(s).
$$
In the case of Poissonian sojourns and Laplacian jumps we can invert this expression and obtain the propagator $p(x,t)$ for different values of $\epsilon$ (see Eqs. (\ref{explicit})-(\ref{explicit_0})). From these expressions we can see some interesting properties due to the existence of correlations such are the transitions from unimodal to bimodal distributions (cf. Fig. \ref{Fig1} and Fig. \ref{Fig2}). 

Although we have been able to solve Eq. (\ref{NMrenewal_2}) in the special case provided by Eq. (\ref{c2}), a general solution to the problem for any form of the joint density $\rho$ seems to be out of reach. However, in many practical situations the degree of dependence between current and past events is weak. In such cases it is possible to derive a perturbation technique which allows for an approximate solution to the above equation to any desired degree of accuracy. We thus write
$$
\rho(\xi',\tau'|\y,\tau)=\rho(\xi',\tau')[1+\epsilon g(\xi',\tau'|\y,\tau)],
$$
where $\rho(\xi',\tau')$ is the unconditional joint density, $g(\xi',\tau'|\y,\tau)$ indicates correlation and $\epsilon$, now a small quantity, measures the strength of such a correlation. The function $g$ is not arbitrary and must obeys some consistency conditions (cf. Eq. (\ref{consistency})) in order to keep the normalization of the $\rho$'s. 

The propagator can be written in the form of an infinite series
$$
\tilde{p}(\omega,s|\y,\tau)=\tilde{q}(\omega,s)+\sum_{n=1}^\infty \epsilon^n \tilde{p}^{(n)}(\omega,s|\y,\tau),
$$
where $\tilde{q}(\omega,s)$ is the propagator when no correlation is present, that is, it corresponds to the propagator of the independent CTRW and is given by Eq. (\ref{solution_0}). The rest of terms $\tilde{p}^{(n)}(\omega,s|\y,\tau)$ ($n=1,2,3,\cdots$) obey the integral equation Eq. (\ref{int_eq_n}) whose solution is given by Eq.~(\ref{recursive_n}), 
what allows us to compute $\tilde{p}^{(n)}(\omega,s|\y,\tau)$ if we know $\tilde{p}^{(n-1)}(\omega,s|\y,\tau)$. Obviously by repeating this operation one can obtain $\tilde{p}^{(n)}(\omega,s|y,\tau)$ for any $n=1,2,3,\cdots$ and, hence, an approximate expression for the propagator to any desired degree of accuracy although, in many cases, the lowest order $n=1$ will suffice. In such a case the explicit expression for the propagator is 
(cf. Eqs. (\ref{tilde_g})-(\ref{first_order}))
$$
\tilde{p}(\omega,s|\y,\tau)=\Biggl\{1+\epsilon\biggl[\tilde{g}_\rho(\omega,s|\y,\tau)+
\frac{\tilde{G}_\rho(\omega,s)}{1-\tilde{\rho}(\omega,s)}\biggr]
+{\rm O}(\epsilon^2)\Biggr\}\frac{\hat{\Psi}(s)}{1-\tilde{\rho}(\omega,s)},
$$
and for the unconditional propagator defined in Eq. (\ref{uncond_def}) we have 
$$
\tilde{p}(\omega,s)=\left[1+\epsilon
\frac{\tilde{G}_\rho(\omega,s)}{1-\tilde{\rho}(\omega,s)}+
{\rm O}(\epsilon^2)\right]\frac{\hat{\Psi}(s)}{1-\tilde{\rho}(\omega,s)}.
$$

We end this work by recalling that our first motivation to treat the problem of dependent CTRW's arose from our work in econophysics. In dealing with extreme time statistics of financial time series, in particular with the mean exit times of the process out of a given interval, we noticed that the observed behavior  cannot be properly described by the traditional ({\it i.e.,} independent) CTRW but one needs some degree of correlation between present and past events \cite{montero_lillo}. In spite of this specific motive we certainly believe that a general development of the dependent CTRW --at least for a Markovian joint density-- may be of broad interest because the independence assumption in the traditional CTRW is just a first approximation for many physical phenomena that are amenable to be studied within the CTRW framework \cite{montroll2,weissllibre}. In any case in forthcoming works we will apply it to financial time series.

\acknowledgments 
The authors acknowledge partial support from Direcci\'on General de Investigaci\'on under contract No. FIS2006-05204.

\appendix

\section{Solution to a recursive integral equation}
\label{Appendix_B}

We will solve the recursive integral equation (\ref{int_eq_n}):
\begin{eqnarray}
\tilde{p}^{(n)}(\omega,s|\y,\tau)&=&Q^{(n-1)}(\omega,s|\y,\tau)\nonumber \\
&+&\int_0^\infty dt' e^{-st'} \int_{-\infty}^{\infty} e^{i\omega x'}\rho(x',t')\tilde{p}^{(n)}(\omega,s|x',t')dx',
\label{b1}
\end{eqnarray}
where $n=1,2,3,\cdots$. To this end we multiply both sides of Eq. (\ref{b1}) by $e^{-s_0\tau+i\omega_0 \y}\rho(\y,\tau)$ and integrate over $\y$ and $\tau$, we obtain
\begin{equation}
\tilde{F}^{(n)}(\omega,s|\omega_0,s_0)=\tilde{Q}^{(n-1)}(\omega,s|\omega_0,s_0)+\tilde{\rho}(\omega_0,s_0)\tilde{F}^{(n)}(\omega,s|\omega,s),
\label{b2}
\end{equation}
where $\tilde{\rho}(\omega_0,s_0)$ is the joint Fourier-Laplace transform of $\rho(\y,\tau)$, 
\begin{equation}
\tilde{F}^{(n)}(\omega,s|\omega_0,s_0)\equiv
\int_0^\infty d\tau e^{-s_0\tau} \int_{-\infty}^{\infty} e^{i\omega_0 \y}\rho(\y,\tau)\tilde{p}^{(n)}(\omega,s|\y,\tau)d\y,
\label{b3}
\end{equation}
and 
\begin{equation}
\tilde{Q}^{(n-1)}(\omega,s|\omega_0,s_0)\equiv
\int_0^\infty d\tau e^{-s_0\tau} \int_{-\infty}^{\infty} e^{i\omega_0 \y}\rho(\y,\tau)Q^{(n-1)}(\omega,s|\y,\tau)d\y.
\label{b3b}
\end{equation}
Setting $\omega_0=\omega$ and $s_0=s$ in Eq. (\ref{b2}) we get
$$
\tilde{F}^{(n)}(\omega,s|\omega,s)=\frac{\tilde{Q}^{(n-1)}(\omega,s|\omega,s)}{1-\tilde{\rho}(\omega,s)},
$$
which introduced back to Eq. (\ref{b2}) yields
\begin{equation}
\tilde{F}^{(n)}(\omega,s|\omega_0,s_0)=\tilde{Q}^{(n-1)}(\omega,s|\omega_0,s_0)+
\frac{\tilde{\rho}(\omega_0,s_0)}{1-\tilde{\rho}(\omega,s)}\tilde{Q}^{(n-1)}(\omega,s|\omega,s).
\label{b4}
\end{equation}
From the definition of $\tilde{F}^{(n)}(\omega,s|\omega_0,s_0)$ given in Eq. (\ref{b3}) we see that the Fourier-Laplace inversion with respect to $\omega_0$ and $s_0$ of this quantity is
$$
\tilde{F}^{(n)}(\omega,s|\omega_0,s_0)\longrightarrow \rho(\y,\tau) \tilde{p}^{(n)}(\omega,s|\y,\tau),
$$
and a similar expression for the inversion of $\tilde{Q}^{(n-1)}(\omega,s|\omega_0,s_0)$. Therefore, the inversion of Eq. (\ref{b4}) reads
\begin{equation}
\tilde{p}^{(n)}(\omega,s|\y,\tau)=\tilde{Q}^{(n-1)}(\omega,s|\y,\tau)+\frac{1}{1-\tilde{\rho}(\omega,s)}\tilde{Q}^{(n-1)}(\omega,s|\omega,s).
\label{b5}
\end{equation}
By combining the definition of $\tilde{Q}^{(n-1)}(\omega,s|\omega,s)$, given in Eq. (\ref{b3b}) when $\omega_0=\omega$ and $s_0=s$, with that of $Q^{(n-1)}(\omega,s|\y,\tau)$ given in Eqs. (\ref{Q^n})-(\ref{Q^0}), we write
\begin{equation}
\tilde{Q}^{(n-1)}(\omega,s|\omega,s)=
\int_0^\infty dt' e^{-st'} \int_{-\infty}^{\infty} e^{i\omega x'}\rho(x',t')G(x',t'|\omega,s)\tilde{p}^{(n-1)}(\omega,s|x',t')dx'
\label{b6}
\end{equation}
where
\begin{equation*}
G(x',t'|\omega,s)\equiv \int_0^\infty d\tau e^{-s\tau}\int_{-\infty}^{\infty}e^{i\omega \y}\rho(\y,\tau)g(x',t'|\y,\tau)d\y.
\end{equation*}
Finally, substituting Eq. (\ref{Q^n}) and Eq. (\ref{b6}) into Eq. (\ref{b5}) we obtain the recursive solution given in Eq. (\ref{recursive_n}).

\end{document}